\documentstyle[aps,prd]{revtex}     
\newcommand{\eff}{{\mathrm{eff}}}
\begin{document} 
 
\title{Comment on "Ponderomotive force due to neutrinos"} 
 
\author{L.O.Silva$^{\P}$, 
 R.Bingham$^{\S}$, J.M.Dawson$^{\P}$, P.K.Shukla$^{\parallel}$, N.L.Tsintsadze$^{\ddag}$, 
and J.T.Mendon\c{c}a$^{\ddag}$ 
 } 
\address{$^{\P}$Department of Physics and Astronomy \\ 
 University of California Los Angeles, Los Angeles CA 90095}
\address{$^{\S}$Rutherford Appleton Laboratory \\
Chilton, Didcot, Oxon OX11 OQX, U.K.}
\address{$^{\parallel}$Institut f\"{u}r Theoretische Physik, Fakult\"{a}t f\"{u}r 
 Physik und Astronomie \\ Ruhr-Universit\"{a}t Bochum, D-44780 Bochum, Germany }
\address{$^{\ddag}$GoLP/Centro de F\'{\i}sica de Plasmas \\
Instituto Superior T\'{e}cnico,
 1096 Lisboa Codex, Portugal}

\draft 
\maketitle 
 
\begin{abstract} 
 The derivation of the ponderomotive force due to neutrinos by Hardy and Melrose
 [Phys.Rev.D {\bf 54}, 6491 (1996)] is based on a 
  flawed analysis of the ponderomotive force concept. Their conclusions also  
 contain an erroneous 
physical assumption related to the neutrino emission in supernovae. 
A correct analysis shows the importance of the ponderomotive 
force due to neutrinos in type II supernovae explosions.
 \end{abstract} 
 
Ref.\cite{hardy} presents several results related with the ponderomotive force of 
the neutrinos, and its role in supernovae explosions. Several claims are in contradiction 
with some of the results presented in their own ref.\cite{hardy} 
and the results already published in the literature. 
Using the methods of Quantum Plasmadynamics, 
 a generalization of Finite Temperature Quantum Field Theory, 
 the mass correction of a single electron due to the presence of a neutrino 
medium is derived (eq.(15) in ref.\cite{hardy}, or eq.(15.\cite{hardy})). 
 This mass correction is equivalent to an effective potential describing the 
effect of a background of neutrinos on single electron dynamics. It describes the 
 electron mass correction due to weak interactions with the neutrinos, and it is the electron counterpart of 
 the effective potential felt by the neutrinos in a background of electrons \cite{bethe}.  
From their eq.(15.\cite{hardy}), it would be straightforward to calculate the force exerted by the 
 neutrino medium 
 on a single electron (which would be just $-\nabla V_\eff$). However, Hardy and Melrose 
first perform a sum over all the electrons in the medium, eq.(16.\cite{hardy}), 
and only then the gradient of the energy density is taken. By taking a sum 
over all the electrons, the total energy density of the medium composed of  
neutrinos {\bf and} electrons is calculated. Therefore the ``ponderomotive force'' 
in eq.(19.\cite{hardy}) is, in fact, 
 the force of the neutrino+electron fluid over some test fluid element
 (interacting via the weak interaction force), and, 
in this sense, it includes the ponderomotive force due to the neutrinos (associated with gradients in the 
 neutrino number 
density) {\bf and} the ponderomotive force due to the electrons 
(associated with gradients in the electron number density).
Therefore, equation (17.\cite{hardy})
 does not describe the force exerted by the neutrino medium over the electrons, as it is claimed. 
 This misinterpretation occurs due to the definition of the ponderomotive force used in \cite{hardy}, 
which can be written as the gradient of the energy density of the neutrino fluid itself (in analogy with the 
approach of Manheimer \cite{manheimer} for the electromagnetic field), 
but cannot be expressed as the gradient of the energy density of the neutrino+electron fluid, 
as represented by Hardy and Melrose \cite{hardy} in eq.(19.\cite{hardy}). 
 
This discrepancy is even more evident from eq.(15.\cite{hardy}), where 
 $\delta m$ is the electron mass correction due to 
a neutrino medium. In this case the effective Hamiltonian for a single electron is written as
\begin{equation}
H_\eff =\sqrt{{\mathbf p}^2+(m_e+\delta m)^2} \simeq {\bf \epsilon}+\frac{m_e}{\epsilon} \delta m
\end{equation} 
 with ${\mathbf p}$ the electron momentum and $m_e$ the electron rest mass, 
and where we have assumed that $|\delta m| \ll m_e$. For the sake of clarity, 
 we neglect anisotropies in the neutrino and antineutrino distribution functions. 
From eqs.(12,13,14.\cite{hardy}), we easily obtain the electron effective Hamiltonian 
\begin{equation}
H_\eff={\bf \epsilon}+\sqrt{2} G_F (n_{\mathrm nu}^+-n_{\mathrm nu}^-)
\end{equation}
where $n_{\mathrm nu}^{+(-)}$ is the neutrino(antineutrino) number density.
Thus the force acting on a single electron, due to the neutrino and antineutrino fluid, is given by 
\begin{equation}
 {\mathbf F}=-\sqrt{2} G_F \nabla (n_{\mathrm nu}^+-n_{\mathrm nu}^-)  \label{eq:pond_f}
\end{equation}
which agrees with other derivations of the ponderomotive force due to neutrinos \cite{silva}.
This result is in clear contradiction with the force derived 
 in ref.\cite{hardy} (eq.(27.\cite{hardy}), when  $n_e^-=0$). According to \cite{hardy},  
 and using eq.(27.\cite{hardy}), the ``ponderomotive force'' exerted on a single electron is 
 ${\mathbf \tilde{F} }/n_e^+$, which is clearly different from eq.(\ref{eq:pond_f}). 
Also, in the treatment of Hardy and Melrose, no 
interaction is assumed between the electrons. Thus, it is physically unreasonable 
 that the ponderomotive force felt by a single electron due to the neutrinos is dependent on the 
 electron density as eq.(27.\cite{hardy}) implies. The correct expression for the 
 ponderomotive force per unit volume 
 due to neutrinos and anti-neutrinos in a background of electrons and 
positrons is then obtained from eq.(\ref{eq:pond_f}), summing the ponderomotive 
 force acting on all the electrons and positrons in the unit volume, thus leading to 
\begin{equation}
 {\mathbf F}_{\mathrm pond}=-\sqrt{2} G_F (n_e^+-n_e^-) \nabla (n_{\mathrm nu}^+-n_{\mathrm nu}^-)
\end{equation}
where $n_e^{+(-)}$ is the electron (positron) number density.

After deriving  the force due to neutrinos, the authors of ref.\cite{hardy} 
apply their results to type II supernovae explosions. In doing so, they discard the 
contribution of the term corresponding to eq.(\ref{eq:pond_f}), 
arguing that ``All species of neutrinos and antineutrinos 
are thought to be produced in the neutrino burst in roughly equal quantities''. However, 
the physical 
scenario in a type II supernovae explosion does not confirm this claim \cite{mayle}. 
Prior to the launching of the shock wave into the outer core, the neutrino signal 
is dominated only by 
electron neutrinos. These electron neutrinos ($\nu_e$) are produced by electron 
capture on protons, both free and 
bound in heavy nuclei. The $\nu_e$ luminosity reaches a peak 
just after the shock wave has moved outside the neutrinosphere ($t \approx 350 \, ms$), 
 as the free protons, produced by shock dissociation of the iron, 
 capture electrons, and a strong $\nu_e$ burst is emitted \cite{mayle}. 
This deleptonization pulse is about 5 ms long and accounts for about 1\%
of the total energy  released in neutrinos of all flavors. 
After this strong deleptonization, neutrinos of all flavors are thermally produced 
by $e^+e^-$ annihilation reactions, and the bulk of the energy is 
emitted in neutrinos of all flavors on neutrino-diffusion timescales ($\approx$ seconds). During the 
strong $\nu_e$ spike, the luminosity of electron antineutrinos 
 ($L_{\bar{\nu}e} \approx 10^{52} erg/s$) is much smaller than the luminosity 
of the electron neutrinos ($L_{\nu e} \approx 4 \times 10^{53} erg/s$) \cite{mayle}.

Hence, it is clear that during the $\nu_e$ burst,  
the component of the ponderomotive force associated with the gradient of the 
neutrino number density plays a dominant role when compared with the component associated with the 
anisotropies of the neutrino distribution function. For typical parameters occurring in a 
supernovae, the ratio between the ponderomotive force $|{\mathrm F_p}|$ (as given by eq.(\ref{eq:pond_f})), 
and the single neutrino-electron collisional force is 
 roughly $|{\mathrm F_p}|/|{\mathrm F_{coll}}| \approx 10^{10} $\cite{bingham2}, 
corresponding to a ponderomotive force with an absolute magnitude of the order of 
$10^{21} \, N m^{-3}$. Since single electron-neutrino scattering already plays a 
significant role in supernovae dynamics \cite{mayle,bethe}, 
it is then clear that ponderomotive force effects should also be included 
in the analysis of supernovae explosions.

During the thermal neutrino emission phase, all types of neutrino flavors are produced, 
 and the role of the ponderomotive force is still important: 
from eq.(\ref{eq:pond_f}) we note that the antineutrinos will push the electrons in the 
opposite direction to the neutrinos but, due to the opposite effective potential affecting the neutrinos and 
antineutrinos \cite{kuo}, 
 the neutrinos will bunch in the regions of lower electron density while the antineutrinos bunch in 
the regions of higher electron density. The ponderomotive force due to neutrinos and 
antineutrinos act together to reinforce the electron 
 density modulations and will still be a fundamental ingredient for this 
instability scenario \cite{bingham3,bingham1}. 
 Recently, another collective mechanism was also considered, relying on neutrino 
scattering by electric field modulations \cite{tsytovich}. 
However, in this process, corresponding to a different physical scenario 
 than the one proposed by Bingham 
{\it et al.} \cite{bingham3,bingham1}, 
 collective effects are much weaker and cannot affect the evolution of the 
exploding star \cite{tsytovich}.

In conclusion, we have shown in this comment the analysis of Hardy and Melrose \cite{hardy} 
of the ponderomotive force due to neutrinos contains some misinterpretations, 
leading to a physically 
unrealistic expression for the force of the neutrinos on the electrons. 
A proper analysis of their formalism gives the correct expression for the ponderomotive force due to neutrinos.
Furthermore, an erroneous assumption about the neutrino and antineutrino species emitted during the 
neutrino burst, led Hardy and Melrose \cite{hardy} to the conclusion of the irrelevance of the ponderomotive force 
during supernovae explosions. 
However, a correct description of the neutrino spectra produced during the neutrino burst \cite{mayle} 
shows that the ponderomotive force due to neutrinos in type II supernovae explosions can impact 
in a significant way the plasma electrons dynamics. During the thermal neutrino emission, the 
ponderomotive force of all flavors still acts as the streaming instability driving mechanism, 
 contributing to the closure of the instability feedback loop.
 Our conclusion is that the ponderomotive force due to neutrinos can 
 play an important role in the explosion mechanism of type II supernovae.


\begin{thebibliography}{99} 
\bibitem{hardy} S.J.Hardy and D.B.Melrose, Phys.Rev. D {\bf 54}, 6491 (1996). 
\bibitem{manheimer} W.M.Manheimer, Phys.Fluids {\bf 28}, 1569 (1985).
\bibitem{bethe}  H.A.Bethe, Phys.Rev.Lett. {\bf 56}, 1306 (1986); 
 L.Wolfenstein, Phys.Rev. D {\bf 17}, 2369 (1978); L.Wolfenstein, Phys.Rev. D {\bf 20}, 2634 (1979).
\bibitem{silva} L.O.Silva, R.Bingham, J.M.Dawson, W.B.Mori, physics/9807049, to appear in 
 Phys.Rev. E (1999); T.Tajima and K.Shibata, {\it Plasma Astrophysics}, (Addison-Wesley, Reading MA, 1997). 
\bibitem{mayle} J.R.Wilson, in {\it Numerical Astrophysics}, eds.J.Centrella, J.Leblanc and R.L.Bowers 
 (Jones and Bartlett, Boston, 1985); R.W.Mayle, J.R.Wilson, and D.N.Schramm, Ap.J. {\bf 318}, 288 (1987); 
J.Cooperstein, Phys.Rep. {\bf 163}, 95 (1988); M.Herant {\it et al}, Ap.J. {\bf 435}, 339 (1994).
\bibitem{bingham2} R.Bingham, H.A.Bethe, J.M.Dawson, P.K.Shukla, J.J.Su, Phys.Lett. A {\bf 220}, 107 (1996).
\bibitem{kuo} T.K.Kuo and J.Pantaleone, Rev.Mod.Phys. {\bf 61}, 937 (1989).
\bibitem{bingham3} L.O.Silva, R.Bingham, J.M.Dawson,  submitted for publication (1998).
\bibitem{bingham1} R.Bingham, H.A.Bethe, J.M.Dawson, J.J.Su, Phys.Lett. A {\bf 193}, 279 (1994).
\bibitem{bethe} H.A.Bethe, ApJ {\bf 412}, 192 (1993).
\bibitem{tsytovich} V.N.Tsytovich, R.Bingham, J.M.Dawson, H.A.Bethe, Astroparticle 
 Physics {\bf 8}, 297 (1998); S.J.Hardy and D.B.Melrose, ApJ {\bf 480}, 705 (1997).
\end{thebibliography}
\end{document}